\documentstyle[prb,aps,epsf,multicol]{revtex}

\tighten
\begin{document}
\draft

\large

\title{
Transmission distribution, $\bbox{{\cal P}(\ln T)}$, of 1D disordered
chain: low-$\bbox{T}$ tail.
}

\author{V. M. Apalkov$^1$ and M. E. Raikh$^2$ \\
$^1$  Department of Physics and Astronomy, Georgia State University,
Atlanta, GA 30303, USA  \\
$^2$Department of Physics, University of Utah, Salt Lake City,
UT 84112, USA }
\maketitle

\begin{abstract}
\large

We demonstrate that the tail of transmission distribution through
1D disordered Anderson chain
is a strong function of the correlation radius of the random potential,
$a$, even when
this radius is much shorter than the de Broglie wavelength, $k_F^{-1}$.
The reason is that the correlation radius defines the phase
volume of the trapping configurations of the random potential, which
are responsible for the low-$T$ tail. To see this, we perform the
averaging over the low-$T$ disorder configurations by first introducing
a finite lattice spacing $\sim a$, and then demonstrating that the
prefactor
in the corresponding functional integral is exponentially small and
depends on $a$ even as
$a \rightarrow 0$.
 Moreover, we demonstrate that
this restriction of the phase volume leads to the dramatic change in the
shape of the tail of ${\cal P}(\ln T)$ from universal Gaussian in
$\ln T $
to a simple exponential (in  $\ln T $) with
exponent depending on $a$. Severity of the phase-volume restriction
affects the shape of the low-$T$ disorder configurations
transforming them from almost periodic (Bragg mirrors) to
periodically-sign-alternating (loose mirrors).

\end{abstract}
\vskip2pc

\begin{multicols}{2}

\vspace{2mm}
\centerline{\bf I. INTRODUCTION}
 
\vspace{5mm}


All the states in one dimension are localized at the
scale of a mean free path, $l_{\epsilon }$. This means
that the {\em typical} value of transmission through
a 1D region of a length, ${\cal L}$, is
$T\sim \exp(-2{\cal L}/l_{\epsilon })$. Since $T$ is
exponentially small, the
subject of recent
theoretical studies\cite{deych,schomerus} is the distribution, ${\cal P}
(\ln T)$, of the log-transmission (and also violation
of the ``orthodox'' 1D localization\cite{berezinskii} for certain
correlated
disorders\cite{izrailev}).
These studies
are mainly focused on the {\em body} of the
distribution ${\cal P} (\ln T)$. A separate issue
is the question about the {\em far tail} of the distribution,
i.e. the behavior of ${\cal P} (\ln T)$ at
$|\ln T| \gg 2{\cal L}/l_{\epsilon }$. This question is directly
related to a more general concept of the anomalously
localized states in disordered
conductors\cite{altshuler,khmelnitskii,falko,mirlin}.
In Ref.\onlinecite{muzykantskii} and in subsequent
paper\cite{smolyarenko}
it was asserted that the small-$T$ tail
is dominated by specific configurations of
the disorder, $V(x)$, namely, the Bragg mirrors.
These configurations are illustrated in Fig.~1.
The potential $V(x)=2{\cal V} \cos (2k_F x)$
opens a gap $2{\cal V}$  centered at
energy $\epsilon = k_F^2/2$. The corresponding
wave function oscillates with a period $\pi/k_F$ and
decays as $\exp\left(-\gamma x\right)$, where
$\gamma = {\cal V}/(2k_F )\ll k_F $
is the decrement. Then we have
$\vert\ln T\vert = 2\gamma {\cal L} = {\cal V}{\cal L}/k_F$.
The important assumption adopted in
Refs. \onlinecite{muzykantskii,smolyarenko}
is that, with exponential accuracy,
${\cal P}\left(\ln T \right)$
can be found
by substituting ${2\cal V}\cos\left(2k_Fx\right)$ into the
``white-noise'' probability,
$\exp\left[-\left\{l_{\epsilon}\int_0^{{\cal L}} dx
V(x)^2/4k_F^2\right\}\right]$, of the fluctuation $V(x)$.
This yields\cite{muzykantskii,smolyarenko}
$\left\vert \mbox{\Large $\ln $} {\cal P}\left(\ln T \right)
\right\vert =
\left(l_{\epsilon}\ln^2 T\right)/2{\cal L}$. Remarkably, the result
coinsides with the asymptote of the ``exact'' solution obtained by
Altshuler and Prigodin\cite{altshuler89} using the Berezinskii
technique\cite{berezinskii}.

The Bragg mirror configurations, $V(x)=2{\cal V} \cos (2k_F x)$,
emerged in Refs. \onlinecite{muzykantskii,smolyarenko} upon
applying the optimal fluctuation approach\cite{halperin,zittartz}.
This approach was specifically
designed to deal with situations when the result is determined
by a particular disorder configuration. The above
log-normal expression for $ {\cal P}\left( T \right)$
corresponds to the saddle point of the functional integral over
disorder configurations. Obviously, the statistical
weight of an ideal Bragg mirror is zero.
Rigorous application of the optimal fluctuation approach
implies taking into account the configurations
close to optimal. This procedure corresponds to
the calculation of the prefactor in the
functional intregral. In most
cases\cite{halperin67,houghton79,brezin,cardy}
the prefactor behaves as a power law and, thus,
cannot compete with the main exponent.

\begin{figure}
\narrowtext
\centerline{
\epsfxsize=3.0in
\epsfbox{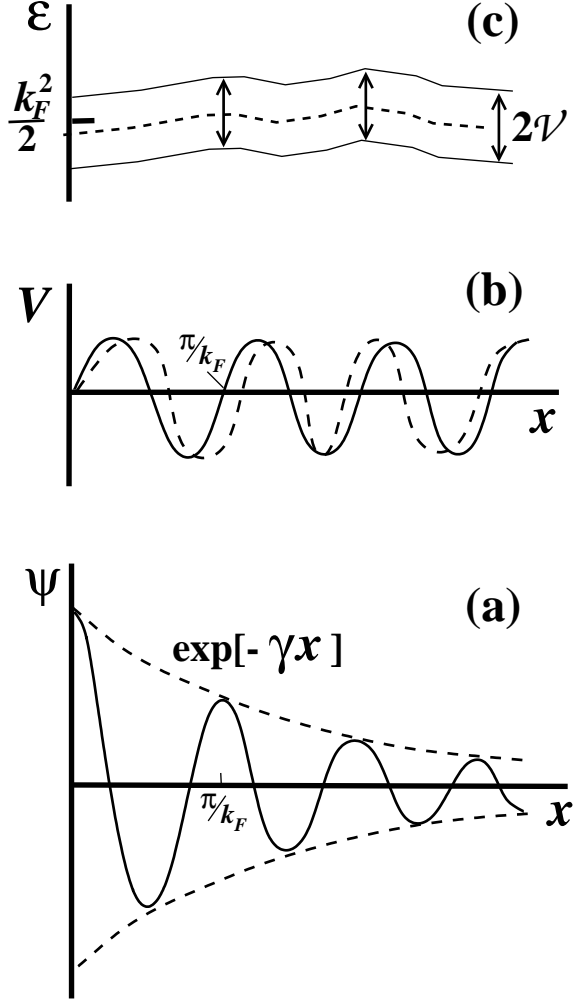}
\protect\vspace*{0.1in}}
\protect\caption[sample]
{\sloppy{
(a) Schematic illustration of the decay of the wave function within
the Bragg mirror.
(b) Solid line: potential fluctuation corresponding to an ``ideal''
Bragg mirror; Dashed line:
``real'' Bragg mirror with fluctuating phase. (c) Fluctuations of
phase result in the fluctuations
of position of the gap center (dashed line) leaving the width of
the gap (solid lines) unchanged.
}}
\label{f1}
\end{figure}

In the present paper we demonstrate that
the situation depicted in Fig.~1a differs drastically
from Refs. \onlinecite{muzykantskii,smolyarenko}
due to a {\em large size} of the optimal fluctuation.
Resulting from this large size, the large
number of ``degrees of freedom'' makes
the prefactor exponentially small, so that,
the final result for $ {\cal P}\left(\ln T
\right)$ is determined by the {\em competition}
of the prefactor and the main exponent.
More specifically, as illustrated in Fig.~1b,c,
weakly perturbed Bragg mirrors include fluctuations with
{\em phase} varying along $x$. These fluctuations are
``dangerous'', in the sense, that they result in
spatial modulation of the gap center (Fig.~1c)
and, thus,  suppress the decrement, $\gamma $.
Large size of a mirror translates into a
large statistical weight of these
dangerous fluctuations, i.e. it severely
restricts the weight of the efficient
Bragg mirrors.

As we demonstrate in the present paper, due to
the reasons listed above,  the
proper application of the optimal fluctuation
approach, i.e. taking prefactor into
account, has dramatic consequences for the shape
of the tail of $ {\cal P}\left(\ln T \right)$. Namely,

\vspace{2mm}
\noindent
(i) The log-normal result\cite{muzykantskii,smolyarenko}
has a ``universal'' form, in the sense, that it contains only
the mean free path, $l_{\epsilon}$.
Thus, it is insensitive to the actual value of
the correlation radius, $a$, of the disorder,
(as long as $a \ll l_{\epsilon}$). In contrast,
we demonstrate that, with prefactor taken into
account, $ {\cal P}\left(\ln T \right)$ depends
on $a$ exponentially strongly even for $a \ll l_{\epsilon}$.

\begin{figure}
\narrowtext
\centerline{
\epsfxsize=3.2in
\epsfbox{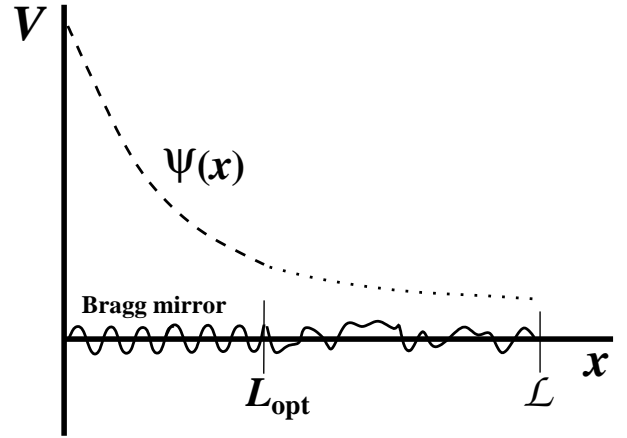}
\protect\vspace*{0.1in}}
\protect\caption[sample]
{\sloppy{Disorder configuration in which the Bragg mirror
occupies only a part $L_{opt}$ of
the total length $\cal L$. The decay of the envelope of
the wave function within the mirror
and in the rest of the sample is illustrated with dashed
and dotted lines, respectively.
}}
\label{f2}
\end{figure}

\noindent
(ii) It was assumed in
Refs.\onlinecite{muzykantskii,smolyarenko} that
the optimal Bragg-mirror fluctuation extends over the
entire region ${\cal L}$. This is indeed the case
for
Gaussian  form of $ {\cal P}\left(\ln T \right)$.
However, with $ {\cal P}\left(\ln T \right)$
having {\em non-Gaussian} form,
it turns out that
the optimal fluctuation corresponds to
the Bragg mirror occupying
{\em only a part}, $L_{opt} < {\cal L}$, of the interval
${\cal L}$, as illustrated in Fig.~2. The underlying
reason for this is that the prefactor makes the Bragg
mirrors very ``costly''.

The paper is organized as follows. In Sec. II we introduce
the discretization, which is always mandatory for the functional
integration. We choose the lattice spacing to be finite,
which is the most convenient discretization for averaging
over disorder configurations of the Bragg-mirror type.
In Sec. III the functional integral, which defines
$ {\cal P}\left(\ln T \right)$, is calculated
{\em with prefactror} in the domain, where the Bragg mirrors
dominate the low-$T$ configurations. In Sec. IV we consider
the low-energy domain, where the Bragg mirrors, being too costly,
become inefficient. We demonstrate that relevant  low-$T$ disorder
configurations in this domain are the {\em loose mirrors}, which
are periodically-sign-alternating on-site energies. As we show in
Sec. IV,  such  loose mirrors form a well-defined subspace in the
space of all possible realizations of the on-site energies. In particular,
they dominate the  functional integral for
 ${\cal P}\left(\ln T \right)$, which we calculate {\em with prefactor}.
In Sec. V we turn to the continuous limit $a\rightarrow 0$. In
contrast to Refs. \onlinecite{muzykantskii,smolyarenko},
we find that, due to the exponentially small prefactor in
the functional integral, it is loose mirrors, {\em extending over a part
of the chain}, rather than the
Bragg mirrors,
{\em occupying the entire chain}\cite{muzykantskii,smolyarenko},
that dominate $ {\cal P}\left(\ln T \right)$ in this limit.
The Sec. VI we trace the reason why the loose mirrors are not
captured in the standard analytical techniques in 1D.

\vspace{5mm}
\centerline{\bf II. GENERAL CONSIDERATIONS}
 
\vspace{5mm}

\centerline{\bf A. Discretization procedure}
 
\vspace{5mm}

To calculate the prefactor of
$ {\cal P}\left(\ln T \right)$, it is necessary, as
in any functional integration, to adopt some sort of
discretization\cite{houghton79}.
In this paper we simply
introduce a finite lattice constant (equal to $1$),
and a finite hopping between the sites (equal to $2$),
so that the problem reduces to 1D Anderson model.
The discrete on-site energies, $V_m$, are random numbers;
their distribution function, $P(V_m)$, has a
characteristic scale $\Delta \ll 1$, which we identify
with r.m.s.
\begin{equation}
 \Delta = \left[ \int _{-\infty}^{\infty } d V_m ~\! V_m^2 P(V_m)
   \right]^{1/2} .
\end{equation}
The discrete version of the ideal ``continuous''
Bragg mirror $V(x)=2{\cal V} \cos (2k_F x)$
has the period $n$ and corresponds to the on-site energies
$V_m=2{\cal V}\cos\left(2\pi m/n\right)$.
Then the discrete analog of the energy $k_F^2/2$ has
the form
\begin{equation}
\epsilon _n = 4 \sin^2\left( \pi/2n \right)   ,
\end{equation}
where the energy is measured from the band edge (equal to
$-2$). To approximate the continuum, unlike Ref. \onlinecite{we},
we will focuse on
the energy interval close to the band edge, i.e. $n\gg 1$.
On the other hand, the energy should be well above the
fluctuation-tail domain, $\epsilon < E_t $, where $E_t$ is
determined from the following consideration.
As follows from the golden rule, for $n\gg 1$,
the mean free path is equal to
$l_{\epsilon}= 8\epsilon_n /\Delta^2$. Then the
conductance, $G_{\epsilon }$, can be written as
$G_{\epsilon } = \epsilon ^{1/2} l_{\epsilon}$.
The upper boundary of the fluctuation-tail domain is
determined by the condition $ G_{E_t } \approx 1$, which
yields $E_t \approx \Delta ^{4/3}$. The fact that we consider
energies above $E_t$ sets the lower bound for the
values of $n$, namely, $n\ll \Delta ^{-2/3}$.

Once the discretization procedure is specified, the averaging
over disorder realizations is well-defined. In particular,
to calculate the statistical weight of the Bragg
mirrors, providing a given value of $\ln T $, one has to
integrate $P(V_m)$ over the deviations of the on-site
energies from $V_m=2{\cal V}\cos\left(2\pi m/n\right)$
with a restriction that the log-transmission
for the set $\{ V_m \}$ is fixed and equal to $\ln T $.
Translating the claim made in
Refs. \onlinecite{muzykantskii,smolyarenko} into
the "discrete" language, this weight is simply equal to
$\prod_m P(2{\cal V}\cos\left[2\pi m/n\right])$, i.e.
the deviations of $V_m$ from
$2{\cal V}\cos\left(2\pi m/n\right)$, that are responsible
for the prefactor, can be neglected
within exponential accuracy.
Below we test this assertion by explicit calculation of
the prefactor. The result of this test
can be summarized as follows.

\vspace{2mm}
\noindent
(i) Weakly disturbed Bragg mirrors (see Fig.~3a) are
indeed the dominating
disorder configurations, providing a given value of
$\ln T $, {\em only} above certain energy,
$E_B \approx \Delta ^{4/5}$, i.e. $n\approx \Delta ^{-2/5}$,
as illustrated in Fig.~4.

\vspace{2mm}
\noindent
(ii) Even for energies bigger
than $E_B$, the prefactor is {\em exponentially}
small. Whether or not it competes with the main
exponent\cite{muzykantskii,smolyarenko} depends
on the length, ${\cal L}$, of disordered region.

\begin{figure}
\narrowtext
\centerline{
\epsfxsize=3.5in
\epsfbox{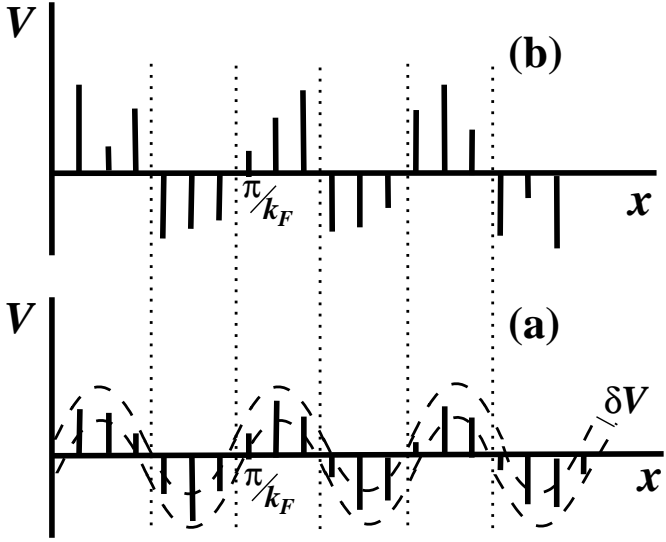}
\protect\vspace*{0.1in}}
\protect\caption[sample]
{\sloppy{(a) Weakly disturbed Bragg mirror on a lattice;
$\delta V$ is the tolerance in the
on-site energies. (b) Loose mirror with ``rigidity'' in
signs of the on-site energies.
}}
\label{f3}
\end{figure}

\begin{figure}
\narrowtext
\centerline{
\epsfxsize=3.1in
\epsfbox{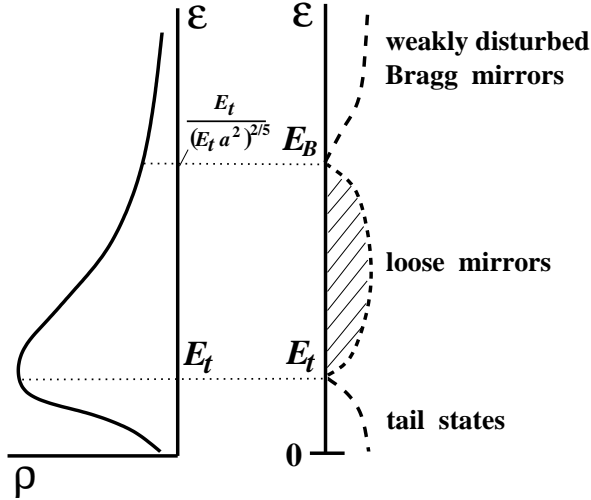}
\protect\vspace*{0.1in}}
\protect\caption[sample]
{\sloppy{Left: schematic plot of the $1D$ density of states
 smeared by disorder. For energies
$\epsilon < E_t$ the states are strongly localized. Right:
low-$T$ disorder configurations
have the form Fig. 3b within the energy domain
$E_t <\epsilon < E_B$ and the form Fig. 3a
for energies $\epsilon > E_B$.
}}
\label{f4}
\end{figure}

\vspace{2mm}
\noindent
(iii)  Our most important finding is that
within a parametrically wide
energy domain, $E_B > \epsilon_n > E_t$, the low-$T$
disorder configurations are dominated by the {\em novel}
entity, which we call ``loose mirrors''. They are
illustrated in Fig.~3b and represent the alternating
regions of equal length $n/2$;  within each region
the values of $V_m$ are either random, but {\em strictly
positive} or random, but {\em strictly negative}.
The ``phase volume'' of these configurations is much
bigger than that of the Bragg mirrors. On the other
hand, for these configurations, at large $n$,
the sign ``rigidity'' of
$V_m$ within each half-period is
sufficient to provide the Bragg reflection.

\vspace{3mm}

\begin{center}
{\bf B. Optimal length of the Bragg mirror
 for a given length of the chain}
\end{center}

\vspace{3mm}

Taking prefactor into account  has a dramatic
effect on the structure of the optimal fluctuation.
To clarify this point,
suppose that the Bragg mirror extends not over
entire distance ${\cal L}$, but only over the
interval $L< {\cal L}$, see Fig.~2.  Denote with $T_L$ the
transmission of the mirror. Then for the
transmission of the entire interval ${\cal L}$ we have
\begin{equation}
\label{sum}
\vert \ln T \vert = \vert\ln T_L \vert +
2\left( \frac{{\cal L}- L}{l_{\epsilon}} \right),
\end{equation}
where the second term describes the transmission through
the region outside the Bragg mirror (Fig.~2).
It is apparent, that the first term in Eq. (\ref{sum})
increases with $L$, whereas the
second term decreases with $L$. This suggests the following
procedure to determine the optimal length of the mirror.
Denote with ${\cal P}_L(\ln T)$ the distribution function
of $T_L$. Then the distribution function of the total
transmission for a given $L$ can be written as
${\cal P}_L\left\{\ln T + 2({\cal L}- L)/l_{\epsilon}\right\}$.
The fact that the Bragg mirror has an optimal length can be
expressed in the form
\begin{eqnarray}
& & \vert \mbox{\Large $\ln$} {\cal P}\left({\cal L},\ln T\right)\vert
             = \nonumber \\
& & ~~~ =\min_{L}\left\vert \mbox{ \Large $\ln$}
 {\cal P}_L\left\{|\ln T| + 2\left( \frac{{\cal L}- L}
{l_{\epsilon}}\right)\right\} \right\vert.
\label{minimum}
\end{eqnarray}
It is seen from Eq. (\ref{minimum}) that the calculation
of the small-$T$ tail of the {\em net} transmission of the entire
interval ${\cal L}$ reduces to the calculation of the function
${\cal P}_L\left(\ln T\right)$, which is the characteristics
of the Bragg mirror only.
If now we use for ${\cal P}_L(\ln T)$
the result  \cite{muzykantskii,smolyarenko}
$~~{\cal P}_L(\ln T)=\exp\left[-\left(l_{\epsilon}\ln^2 T\right)/2L
\right] $, which was obtained neglecting the prefactor,
then the minimum in Eq.~(\ref{minimum}) would
correspond to $L = {\cal L}$, i.e. to the Bragg mirror extending
over the {\em entire interval} ${\cal L}$.
Below we demonstrate that, once the prefactor is taken
into account, the true optimal fluctuation corresponds to
a ``short'' Bragg mirror, $L < {\cal L}$, within a parametrically
wide interval of $\left| \ln T \right|$.

\vspace{3mm}

\begin{center}
{\bf III. WEAKLY DISTORTED BRAGG MIRRORS}
\end{center}

\begin{center}
{\bf A. Calculation of the functional integral}
\end{center}

\vspace{2mm}

Here we consider the case when the deviations, $\delta V_m$, of the
on-site energies, $V_m$, from the optimal values,
$V_m=2{\cal V}\cos\left(2\pi m/n\right)$,  are relatively small.
For concreteness we choose the Gaussian distribution of the on-site
energies, $P(V) = \pi^{-1/2}\Delta ^{-1} \exp (-V^2/\Delta^2)$.
To calculate the prefactor due to  small deviations, $\delta V_m$,
we adopt the assumption that  $\delta V_m$ are {\em homogeneously}
distributed within a small interval (tolerance) $\delta V \ll \Delta $
(Fig.~3a).
On the one hand, this assumption leads to a drastic simplification
of the calculation. On the other hand, as we will see below,
it yields an asymptotically correct result.

With homogeneously distributed $\delta V_m$, the
statistical weight of distorted Bragg mirror,
${\cal P}_L$, can be easily expressed through the
tolerance $\delta V$
\begin{eqnarray}
{\cal P}_L & = & \left(\frac{\delta V}{\Delta}\right)^L
\exp\left[-\left(\frac{1}{\Delta^2}\right)\sum_m V_m^2\right] \nonumber \\
 & = & \exp\left[-L\ln\left(\frac{\Delta}{\delta V}\right) -
\frac{{\cal V}^2L}{2\Delta^2}\right].
 \label{simplified}
\end{eqnarray}
In Eq.~(\ref{simplified}) we have  assumed that
$\delta V$  not only smaller than $\Delta$,
but even stronger condition $\delta V \ll \Delta^2/{\cal V}$ is met.
We will check this condition below.

We now incorporate the fluctuations
$\delta V_m$ into  the log-transmission of the
Bragg mirror, $\ln T$.
 As it was pointed out above, random shifts, $\delta \epsilon_i$,
of the gap center
reduce the decrement $\gamma = {\cal V}/2k_F = {\cal V}n/2\pi$
within each period. This is due to the local detuning from
the Bragg resonance. Quantitatively, the reduction of the
decrement, $\gamma $, can be expressed as
\begin{equation}
\label{gamma}
\gamma \left( \delta \epsilon _i\right) = \gamma
\sqrt{1- \left(\frac{\delta \epsilon_i }{{\cal V}}\right)^2}.
\end{equation}
As a result, instead of $2\gamma L$ in the absence of fluctuations,
the expression for $\vert\ln T \vert $ modifies to
\begin{eqnarray}
\vert\ln T \vert & = & 2n \sum _i \gamma (\delta \epsilon _i)
  \nonumber \\
& \approx &
 2 \gamma L - n \gamma \sum _i \left( \frac{\delta \epsilon _i}{{\cal V}}\right)^2 .
\label{reduced}
\end{eqnarray}
Consider now a given period, $i$, containing $n$ sites.
Denote with $V^{(i)}_{m}$ the on-site energies within
this period. Then the shift, $\delta \epsilon _i$,
of the gap center for this period can be expressed through
$V^{(i)}_{m}$ via a discrete Fourier transform as follows
\begin{equation}
\label{d_e_0}
\delta \epsilon _i=
\left(\frac{\pi}{n}\right)^2
\frac{ \sum_{m=1}^{n} V_m^{(i)} \sin \left( \frac{\pi m}{n}\right)  }
{ \sum_{m=1}^{n} V_m^{(i)} \cos \left( \frac{\pi m}{n}\right) },
\end{equation}
where the summation is performed over the sites within the
$i$-th period. Obviously, for an ideal Bragg mirror,
$V_m=2{\cal V}\cos\left(2\pi m/n\right)$, we obtain from
Eq.~(\ref{d_e_0}) that $\delta \epsilon _i=0$. In the presence of
fluctuations, $\delta V_m$, the typical
value of $\delta \epsilon _i$ is proportional to $ \delta V $
and can be estimated from Eq.~(\ref{d_e_0}) as follows.
The numerator is the sum of $n$ random numbers, each being
$\sim \delta V$. Thus, the typical value of the numerator
is $n^{1/2}\delta V$. On the other hand,
the denominator is equal to $n {\cal V}/2$. Then we obtain
\begin{equation}
\label{d_e_1}
\delta \epsilon _i =\frac{C\epsilon_{2n} }{(2n)^{1/2}}
\left(\frac{\delta V}{{\cal V}}\right) =
\frac{\pi^2 C}{2^{1/2}n^{5/2}}
\left(\frac{\delta V}{{\cal V}}\right),
\end{equation}
where the constant $C$ is of the order of $1$.

Looking at Eq. (\ref{gamma}), it might seem that the condition
$\delta V \ll {\cal V}$
of the  weak distortion of the Bragg mirror by fluctuations,
and the condition $\delta \epsilon_i \ll {\cal V}$
of the weak reduction of the decrement are quite different.
It turns out, as we will see later, that $\delta V \ll {\cal V}$ insures
that $\vert \gamma_i - \gamma \vert \ll \gamma$, and thus justifies
the expansion of $\gamma (\delta \epsilon _i)$
used in Eq.~(\ref{reduced}). Substituting Eq.~(\ref{d_e_1}) into
 Eq.~(\ref{reduced}) we get
\begin{eqnarray}
\vert\ln T \vert  - \frac{{\cal V}L n}{\pi}  & \approx &
- C^2 \frac{{\cal V}L}{2\pi}
\left(\frac{ \epsilon_{2n}\delta V}{{\cal V}^2}\right)^2 \nonumber \\
 & \approx & -8 \pi^3 C^2
\left(\frac{ L \delta V ^2}{n^4 {\cal V}^3}\right) .
\label{expansion2}
\end{eqnarray}
Using the fact that the r.h.s. in Eq. (\ref{expansion2}) is much smaller
than $\vert \ln T \vert$, we can express ${\cal V}$ through
$\vert \ln T \vert$ as follows
\begin{equation}
\label{V}
{\cal V} = \frac{\pi \vert\ln T \vert }{ n L } +
   8\pi C^2 \frac{L^3 \delta V^2 }{ n^2 \vert\ln^3 T \vert } .
\end{equation}
Further steps are straightforward. Using Eq. (\ref{V}), we can rewrite
the exponent in Eq. (\ref{simplified}) as
\begin{eqnarray}
& & |\ln {\cal P}_L | = \left\{ L\ln\left(\frac{\Delta}{\delta V}\right) +
\frac{{\cal V}^2L}{2\Delta^2} \right\}
   \nonumber \\
&  &  =
\frac{\pi^2 \ln^2 T}{2n^2 \Delta ^2 L} \! + \!
\left\{
  L\ln\left(\frac{\Delta}{\delta V}\right) \! + \!
\frac{8\pi^2C^2 L^3 \delta V^2 }{n^3 \Delta ^2 \ln^2 T}
  \right\} .
\label{exponent}
\end{eqnarray}
Now it is easy to see that there exists the optimal
tolerance
\begin{equation}
\label{V_opt}
\delta V = \delta V_{opt}  = \frac{{\cal V}\Delta n^{5/2}}{4\pi^2}=
\frac{n^{3/2} \Delta \vert \ln T \vert}{4\pi C L},
\end{equation}
for which $\vert \ln {\cal P}_L \vert$ is minimal and is equal to
\begin{equation}
\label{optimal}
\left\vert\ln{\cal P}_L\right\vert =
    \frac{\ln^2 T}{2 \Delta ^2 L} \left( \frac{\pi}{n}\right)^2
+L \Lambda\left(T\right),
 \end{equation}
where
\begin{equation}
\label{Lambda}
\Lambda\left(T\right)= \ln\left(\frac{\Delta}{\delta V_{opt}}\right)=
\ln\left(\frac{4\pi CL}{n^{3/2}| \ln T |}\right)
\end{equation}
depends on $T$ very weakly. It is also seen from Eq. (\ref{Lambda})
that the $C$ enters into the final
result only as a factor under the logarithm, so that our assumption
about homogeneous distribution of $\delta V_m$ is justified.

\end{multicols}
\widetext
\vspace*{-0.2truein} \noindent  \hrulefill \hspace*{3.6truein}

Now, in order to calculate the tail of the transmission distribution,
${\cal P}\left({\cal L}, \ln T\right)$,  we substitute
Eq.~(\ref{optimal}) into Eq.~(\ref{minimum})
\begin{equation}
\label{optimal_1}
\left\vert\ln{\cal P}_L\right\vert =  \min _{L} \left\{  \frac{1}{2L}
\left( \frac{\pi}{n \Delta }\right)^2
\left[ |\ln T| + 2\left( \frac{{\cal L}- L}
{l_{\epsilon}}\right)  \right]^2
+L \Lambda\left(T\right)   \right\}.
 \end{equation}
Next we perform
minimization with respect to $L$. This yields the following equation of the optimal
$L=L_{opt}$
\begin{equation}
\label{Lopt}
L _{opt}= \frac{ \pi |\ln T| }{\sqrt{2 \Lambda } n \Delta }
   \left[1 +2 \left( \frac{{\cal L}- L_{opt}} {l_{\epsilon} |\ln T|}  \right) \right]
\left[  1 - \frac{2}{\Lambda }\left(\frac{\pi }{n \Delta }  \right)^2
\frac{|\ln T|}{l_{\epsilon }L_{opt}}  \left\{ 1 +2 \left(
\frac{{\cal L}- L_{opt}} {l_{\epsilon} |\ln T|}  \right) \right\}   \right]^{-1/2}.
\end{equation}
Since we are interested in anomalously low transmissions, $|\ln T |\gg {\cal L}/l_{\epsilon}$,
the second term in the first square bracket in Eq. (\ref{Lopt}) is small.  The second term
in the second square bracket contains an additional parameter
$\sim |\ln T|/L_{opt}(n^2\Delta^2l_{\epsilon})$. Since $l_{\epsilon} =8\epsilon_n/\Delta^2$,
the combination $n^2\Delta^2l_{\epsilon}$ is $\sim 1$. Thus, the above parameter reduces to
$|\ln T|/L_{opt}$, which is also small. More precisely, it is of the order of $l_{\epsilon}^{-1/2}$.
Neglecting second terms in both square brackets, and substituting $L_{opt}$ from Eq. (\ref{Lopt})
into Eq. (\ref{optimal_1}), we arrive at the final result
\begin{eqnarray}
\left\vert \mbox{\Large $\ln$ } \!
                           {\cal P}\left(\ln T, {\cal L}\right)
       \right\vert = & &
  \left(\frac{\pi \sqrt{2\Lambda }}{n \Delta } \right)
                                       \vert \ln T \vert
~,~~~~~~~~~~~~~~~~~~  \left(\frac{\sqrt{2\Lambda}}{\pi}\right)
                           n \Delta   {\cal L}   >
         \vert \ln T \vert > n^2 \Delta ^2{\cal L}
\label{small}
    \\
 \left\vert \mbox{\Large $\ln$ } \!
                          {\cal P}\left(\ln T, {\cal L}\right)
\right\vert     = & &  \frac{1 }{2 {\cal L}}
 \left( \frac{\pi}{n\Delta }\right)^2  \ln^2 T
+{\cal L}   \Lambda \left(T\right)    ~,~~~~~
 \vert \ln T \vert > \left(\frac{\sqrt{2\Lambda}}{\pi}\right)
                           n \Delta {\cal L}.
\label{large}
\end{eqnarray}
It is instructive to rewrite the above result in terms of
energy, $\epsilon _n \approx \left( \pi/n\right)^2 $,
and conductance, $G_n \approx 1/(n^{3}\Delta ^{2})$
\begin{eqnarray}
\left\vert \mbox{\Large $\ln$ } \!
                           {\cal P}\left(\ln T, {\cal L}\right)
       \right\vert = & &
  \left(\frac{2 \pi^3 \Lambda G_n}{\sqrt{ \epsilon_n } } \right)^{1/2}
                                       \vert \ln T \vert
~,~~~~~~~~~~~~~~~~~~
\left(\frac{2 \Lambda \sqrt{ \epsilon_n } }{\pi ^{3} G_n }\right)^{1/2}
                             {\cal L}   >
         \vert \ln T \vert >
  \left(\frac{ \sqrt{\epsilon_n }}{\pi G_n}  \right) {\cal L}
\label{small_E}
    \\
 \left\vert \mbox{\Large $\ln$ } \!
                          {\cal P}\left(\ln T, {\cal L}\right)
\right\vert     = & &  \frac{\pi^3 }{2 {\cal L}}
 \left( \frac{G_n }{ \sqrt{\epsilon_n } }\right)  \ln^2 T
+{\cal L}   \Lambda \left(T\right)    ~,~~~~~
 \vert \ln T \vert >
\left(\frac{2 \Lambda \sqrt{ \epsilon_n } }{\pi ^{3} G_n }\right)^{1/2}
                             {\cal L} .
\label{large_E}
\end{eqnarray}

\hspace*{3.6truein}\noindent  \hrulefill

\begin{multicols}{2}

\begin{figure}
\narrowtext
\centerline{
\epsfxsize=3.2in
\epsfbox{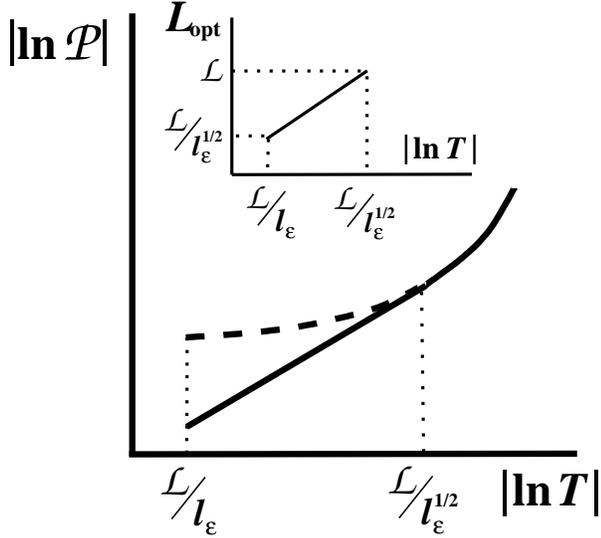}
\protect\vspace*{0.1in}}
\protect\caption[sample]
{\sloppy{Solid line: our main result Eqs. (\ref{small}), (\ref{large})
 for the low-$T$ tail of the transmission distribution. Dashed line:
log-normal ${\cal P}(T)$
of Refs. \onlinecite{muzykantskii,smolyarenko}. Inset: the portion
of the chain occupied with
Bragg (loose) mirror.
}}
\label{f5}
\end{figure}

The behavior Eqs. (\ref{small}), (\ref{large}) is illustrated
schematically in Fig.~5. We see, that the log-normal dependence of
Refs.~\onlinecite{muzykantskii,smolyarenko} takes place only for
very small transmission coefficients
$\vert \ln T \vert > {\cal L}/\sqrt{l_{\epsilon}}$. Only in this domain
the Bragg mirror extends over the entire interval,
and the prefactor [second term in Eq. (\ref{large})]
is smaller than the main exponent.
Within a wide domain
${\cal L}/\sqrt{l_{\epsilon}} > \vert \ln T \vert >
{\cal L}/l_{\epsilon}$
the probability, ${\cal P}\left({\cal L}, \ln T\right)$, behaves as
a {\em simple exponent}. The underlying reason for such a behavior
is that the dependences of the main exponent and of the prefactor
on $L$ are {\em opposite}. As a result, there exists an optimal mirror
length
\begin{equation}
\label{L_opt}
L=L_{opt}=\frac{\pi\left\vert\ln T\right\vert }{\sqrt{2\Lambda}n\Delta},
\end{equation}
which leads to the result Eq.~(\ref{small}).
 Simple exponent Eq. (\ref{small}) corresponds to the
situation  when $L_{opt}$ is shorter than the length of the interval,
${\cal L}$. More precisely, the portion of the interval ${\cal L}$,
occupied by the Bragg mirror, is given by $L_{opt}/{\cal L} =
\pi\left\vert\ln T\right\vert/\sqrt{2\Lambda}n\Delta{\cal L}$.
Within the domain ${\cal L}/\sqrt{l_{\epsilon}} > \vert \ln T \vert >
{\cal L}/l_{\epsilon}$ this portion changes from $1$ to a small
value $\left(\Delta /G \right)^{1/3}\ll 1$, as illustrated in the
inset in Fig.~5.


\vspace{2mm}

\begin{center}
{\bf B. Justifications of the assumptions}
\end{center}

\vspace{2mm}

The above calculation was based on three assumptions

\vspace{2mm}
\noindent(i) $\delta V \ll \Delta^2/{\cal V}$; we used this
condition in the expression Eq.~(\ref{simplified}) for
the probability, ${\cal P}_L$.

\vspace{2mm}
\noindent(ii) $\delta V \ll {\cal V}$; this is the condition
that the Bragg mirror is well defined. It was also
used in deriving Eq.~(\ref{simplified}).

\vspace{2mm}
\noindent(iii) $\delta \epsilon _i \ll {\cal V}$; this
condition was used in expansion Eq.~(\ref{reduced}).

From the result Eq.~(\ref{V_opt}) of the above
calculation we find $\delta V _{opt} /{\cal V} = \Delta n^{5/2}/
(4 \pi ^2)$. Thus, the assumption (ii) is valid under
the condition
\begin{equation}
\label{condition}
\Delta n^{5/2} \ll 1.
\end{equation}
Below we demonstrate that the same condition Eq.~(\ref{condition})
guarantees the validity of the other two assumptions.

\vspace{2mm}
\noindent {\em Assumption} (i).  Within the domain
${\cal L}/\sqrt{l_{\epsilon}} > \vert \ln T \vert >
{\cal L}/l_{\epsilon}$, where ${\cal P}\left({\cal L}, \ln T\right)$
behaves as a simple exponent, the amplitude of the Bragg mirror,
given by ${\cal V} = \pi |\ln T| /nL_{opt}$, is equal to
${\cal V} = \sqrt{2\Lambda}\Delta$ and does not depend on
${\cal L}$, where, as follows from Eq.~(\ref{Lambda}),
$\Lambda = \Lambda \left(L_{opt}\right)=
\ln\left(4\pi^2 C /\sqrt{2\Lambda } \Delta n^{5/2}\right)
\approx \ln\left(1/\Delta n^{5/2}\right) $.
To check the assumption (i) we rewrite the
ratio ${\cal V} \delta V / \Delta^2 $ as
$ (\delta V/{\cal V}) ({\cal V}^2 / \Delta^2)=
(\Delta n^{5/2}/4 \pi^2 ) ({\cal V}^2 / \Delta^2) =
\frac{1}{2 \pi^2} \Delta n^{5/2}  \ln\left(1/\Delta n^{5/2}\right)$.
We see that  ${\cal V}  \ll  \Delta^2 /\delta V$ holds under the
condition $\Delta n^{5/2} \ll 1$, which is precisely the
condition (\ref{condition}).

\vspace{2mm}
\noindent {\em Assumption} (iii)
From Eq.~(\ref{d_e_1}) we have the following estimate for the
ratio $\delta \epsilon _i /{\cal V} \ll 1$
\begin{equation}
\label{estimate}
\frac{ \delta \epsilon _i}{{\cal V}}  =
\frac{\pi^2 C}{2^{1/2}n^{5/2}}
\left(\frac{\delta V}{{\cal V}^2}\right) .
\end{equation}
Substituting into this equation the optimal value
$\delta V _{opt} = {\cal V}\Delta n^{5/2} /(4 \pi ^2)$,
we obtain $\delta \epsilon /{\cal V} \sim \Delta /{\cal V} =
[2 \Lambda (L_{opt})]^{-1/2}$. On the other hand,
$\Lambda \left(L_{opt}\right) =
\ln\left(1/\Delta n^{5/2}\right)$ is large under the same
condition Eq.~(\ref{condition}). This large
logarithm justifies the assumption (iii).

In conclusion of this Section we would like to make the
following two remarks:

\vspace{2mm}
\noindent(1)  The expression for the decrement
$\gamma = {\cal V }n/2\pi$ is the result of the two-wave
approximation, within which propagating and Bragg-reflected
waves are coupled only in the first order, i.e. by a
single harmonics of periodic potential. For two-wave
approximation to be valid, the second-order coupling
matrix elements must be much smaller than ${\cal V}$.
The estimate for these second-order elements is
$\sim {\cal V}^2/\epsilon_n \sim {\cal V}^2 n^2 $.
Thus, the two-wave approximation is valid if
${\cal V}^2n^2 \ll {\cal V}$, i.e. ${\cal V} n^2 \ll 1$.
As it is seen from Eq. (\ref{condition}),
${\cal V} \ll \Delta ^{1/2} n^{-5/4}$. Thus
${\cal V} n^2  \ll \left[ \Delta ^{1/2} n^{-5/4}\right]
n^2 = \left[ \Delta  n^{5/2}\right]^{1/2}/n^{1/2} $.
We see that the applicability condition of the
two-wave approximation is weaker than the
main condition $\Delta  n^{5/2}\ll 1$.

\vspace{2mm}
\noindent(2) The condition Eq.~(\ref{condition}) implies
that the energy $\epsilon_{n}$ exceeds $\Delta^{4/5}$. This, in turn,
suggests that the conductance $G_n = k_Fl_{\epsilon}$ for
$\epsilon = \epsilon_{n}$ is equal to
$G_n = \left(\Delta^2n^3\right)^{-1}$, and is large by virtue of this
condition.

\vspace{3mm}

\centerline{\bf IV. ``LOOSE'' MIRRORS}

\vspace{3mm}

\centerline{\bf A. Density of the loose mirrors}

\vspace{3mm}

We now turn to the case of low energies. More precisely,
we consider the domain $E_B > \epsilon_n > E_t$ (Fig.~4).
The upper boundary of this domain corresponds to
$\Delta n^{5/2} \approx 1$, whereas the lower boundary
corresponds to $\Delta n^{3/2} \approx 1$. For energies
$\epsilon _n > E_B$ the transmission is dominated by
weakly disturbed Bragg mirrors, as discussed in the
previous Section. For energies $\epsilon < E_t$ we have
$G_{\epsilon } < 1 $, i.e. these energies correspond to the
tail states.

As we enter the low energy (large-$n$) domain,
the key component of the above scenario of weakly
disturbed mirrors gets violated. Namely,
at $n \sim \Delta^{-2/5}$ we have $\delta V \sim {\cal V}$.
This implies that almost sinusoidal
Bragg mirror cannot retain its role as an optimal fluctuation,
which is responsible for low-$T$ values.


In general, optimal fluctuation constitutes a saddle-point in
the functional space. In the previous Section, by demonstrating
that the disorder configurations, contributing to the
functional integral, differ weakly (by $\delta V \ll {\cal V}$)
from the optimal configuration, we have justified that the
saddle point is well defined, or, in other words, the
expansion around the saddle point yields a narrow width of
the Gaussian in the functional space. In this Section we
demonstrate that in the energy domain
$E_B > \epsilon_n > E_t$ there exists a well-defined subspace
of all realizations of the on-site energies, $\{ V_m\}$,
which assumes the role of a saddle point. We dub
the elements of this subspace as ``loose'' mirrors.
A loose mirror is a configuration of alternating regions
of $n$ {\em random}, but {\em positive} $V_{m}$ and
 $n$ {\em random}, but {\em negative} $V_{m}$. It
is illustrated in Fig.~3.
Obviously, the statistical weight of the loose mirrors
is small. It is easy to see that this weight is
equal to $2^{-L}$. Most importantly, despite the
randomness of $V_m$, the fact of the {\em sign rigidity} within
each interval of length, $n$, is sufficient for the formation
of the Bragg gap with the {\em well-defined width}, and
thus, for generating the low transmission coefficients.

The key element of calculation of ${\cal P} (\ln T)$
in the regime of weakly distorted Bragg mirrors was the
expansion Eq.~(\ref{reduced}), which expressed the fact
that the decrement $\gamma $ weakly fluctuates from
period to period. It turns out that in the
regime of loose mirrors, $ \Delta n^{5/2} \gg 1$,
these fluctuations are also weak. This can be seen
from Eq.~(\ref{estimate}). Since in the regime of
loose mirrors the only scale for ${\cal V}$ and
$\delta V $ is $\Delta $, Eq.~(\ref{estimate}) yields
$\delta \epsilon _i/{\cal V}
\sim (\Delta n^{5/2})^{-1} $, which is small in the
regime of loose mirrors.
Thus, the expansion Eq.~(\ref{reduced}) is applicable
in this regime as well.

To calculate the distribution ${\cal P} (\ln T)$ in the
regime of loose mirrors, the calculation in the
previous Section should be modified in the following
way. For loose mirrors the ``period'' consists of
interval of $n$ positive $V_m$ followed by an
interval of $n$ negative $V_m$. The magnitude of the
gap $2{\cal V}$ and corresponding decrement,
$\gamma = n{\cal V}/2\pi $,
are determined by discrete Fourier
component of this realization of the on-site energies.
Then the expansion analogous to Eq.~(\ref{expansion2})
takes the form
\begin{equation}
\label{expansion_loose}
\vert\ln T \vert \approx \frac{{\cal V}L n}{\pi}.
\end{equation}
Then the corresponding expression for ${\cal V}$,
analogous to Eq.~(\ref{V}), reads
\begin{equation}
\label{V_loose}
{\cal V} = \frac{\pi \vert\ln T \vert }{ n L } .
\end{equation}
It is obvious that the typical value of ${\cal V}$ is
$\sim \Delta $ with variance is $\sim \Delta /n^{1/2}$.
It can be demonstrated that  the full distribution
of ${\cal V}$ is given by
\begin{equation}
\label{pV}
p({\cal V}) =  \left(\frac{n}{\pi \beta_2 \Delta^2 }\right)^{1/2}
\!\!\! \exp\left[ -
\frac{n({\cal V}-\beta_1 \Delta )^2}{\beta_2 \Delta ^2}  \right],
\end{equation}
where the constants $\beta_1$ and $\beta_2$ depend on the actual
distribution $P(V)$. For Gaussian $P(V)$ they assume the values
$\beta_1 = 4\pi^{-3/2}$ and $\beta_2= 1/2-1/\pi$.

Analogously to Eq.~(\ref{simplified}), the actual calculations
reduces to optimization with respect to $L$ of the product
\begin{eqnarray}
{\cal P}_L & = & \left(\frac{1}{2}\right)^L  \left[ ~\! p ~\! ({\cal V})
 ~\! \right]^{L/n} =\nonumber \\
&  = &  \left(\frac{1}{2}\right)^L
\left[ p\left(
\frac{\pi \vert\ln T \vert }{ n L }
\right)  \right]^{L/n}  \!\! ,
\label{P_loose}
\end{eqnarray}
where the power $L/n$ emerges from the product over periods.
With Gaussian $p({\cal V})$, given by Eq.~(\ref{pV}), this optimization
can be performed analytically in a similar fashion as in the
previous Section, yielding
\begin{equation}
\label{L_loose}
L_{opt} = \frac{ \pi | \ln T |}{
 n \Delta  \beta _0^{1/2}},
\end{equation}
where $\beta_0 = \beta_2 \ln 2 + \beta_1^2$  is
the constant of the order of 1.
The corresponding value of the gap width is
\begin{equation}
\label{V_loose2}
{\cal V} _{opt} = \frac{ \pi | \ln T |}{
 n L_{opt}} = \Delta \beta _0^{1/2} .
\end{equation}

\end{multicols}
\widetext
\vspace*{-1mm} \noindent  \hrulefill \hspace*{3.6truein}

The result Eq.~(\ref{L_loose}) is quite similar to Eq.~(\ref{L_opt}),
and differs only by replacement of the logarithmic factor,
$2\Lambda $,
by a constant $\beta_0 $, which is of the order of 1.
Correspondingly, the final results for ${\cal P} (\ln T, {\cal L})$
are quite similar to Eqs.~(\ref{small_E})-(\ref{large_E})
\begin{eqnarray}
\left\vert \mbox{\Large $\ln$ } \! {\cal P}\left(\ln T, {\cal L}\right)
       \right\vert = & & \beta _{eff}
\left(  \frac{2 \pi^3 G_n}{ \sqrt{\epsilon_n}} \right)^{1/2}
                                       \vert \ln T \vert
,~~~~~~~~~~ \frac{{\cal L}}{\pi}
\left(  \frac{ \beta_0\sqrt{\epsilon_n}}{\pi G_n} \right)^{1/2}  >
         \vert \ln T \vert >
\left(\frac{\sqrt{\epsilon_n}}{\pi G_n} \right) {\cal L}
\label{small1}
    \\
\left\vert \mbox{\Large $\ln$ } \! {\cal P}\left(\ln T, {\cal L}\right)
       \right\vert = & &
\left( \frac{\pi^3 G_n}{\beta_2 \epsilon _n^{1/2} {\cal L}} \right)
  \ln ^2 T -
\left(  \frac{2 \pi^3 \beta_1^2 G_n}{\beta_2^2 \sqrt{\epsilon_n}}
\right)^{1/2} \!\!\! | \ln T| -
\left( \frac{\beta_0}{\beta_2} \right) {\cal L}
     ,~~~~
 \vert \ln T \vert >
\frac{{\cal L}}{\pi}
\left(  \frac{\beta_0 \sqrt{\epsilon_n}}{\pi G_n} \right)^{1/2}  \!\! ,
\label{large1}
\end{eqnarray}
where $\beta _{eff}=\left[  \beta_0^{1/2}  - \beta_1 \right]/\beta _2 $.
For Gaussian distribution of the on-site energies we have $\beta_{eff}\approx 0.46$.
The results Eqs.~(\ref{small1}), (\ref{large1}) were obtained
assuming that loose mirrors are well-defined entities, in  the
sense, that the subspace that they constitute within
the functional space has a sharp boundary. In the next
subsection we examine
the width of this boundary and demonstrate that this width is
indeed relatively small.

\hspace*{3.6truein}\noindent  \hrulefill

\begin{multicols}{2}

\vspace{2mm}

\centerline{\bf B. Tolerance of the loose mirrors}

\vspace{2mm}
In order to examine to what extent the loose mirrors are
well defined,
we consider below two generic sources of violation of the sign
rigidity, which are illustrated in Fig. 6.

\vspace{2mm}
\noindent(i)
We allow the on-site energies within ``positive'' periods
to assume slightly negative values, restricted by $-W$ (see
Fig.~6a), and
the on-site energies within ``negative'' periods to assume small
positive values, restricted by $W \ll \Delta $.
This allowance increases
exponentially the number of configurations constituting
the loose mirrors. On the other hand, such an allowance
suppresses the gap. As a result of these competing trends,
there exists an optimal value of $W$, that maximizes
${\cal P}(\ln T, {\cal L})$.

\vspace{2mm}
\noindent(ii) We allow a {\em small} portion, $\kappa $, of
on-site energies to assume the ``wrong'' sign preserving
their magnitude $\sim \Delta $. This allowance also
increases the number of loose mirrors and suppresses
the gap. Thus there exists an optimal $\kappa \ll 1$,
which we calculate below.

The quantitative characteristics of the "quality" of the
loose mirror is the fluctuation, $\delta \epsilon $, of the
gap center due to the above violations, which is analogous to
the tolerance $\delta V$ of Bragg mirror in the previous Section.

\begin{figure}
\narrowtext
\centerline{
\epsfxsize=3.2in
\epsfbox{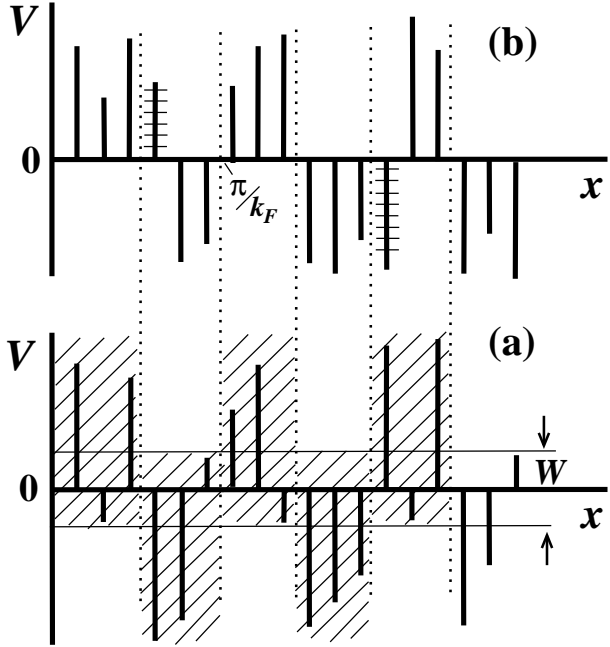}
\protect\vspace*{0.1in}}
\protect\caption[sample]
{\sloppy{Possible violations of the  sign rigidity of a loose mirror:
(a) small ($\sim  W  \ll \Delta$)
 on-site energies with ``wrong'' sign are allowed;   (b) sparse
``large'' ($\sim \Delta$)  on-site energies having
the ``wrong'' sign (hash-marked lines) are allowed.
}}
\label{f6}
\end{figure}

The enhancement of the portion of the loose mirrors
due to allowance, $W$, can be estimated as $(1+W/\Delta )^{L_{opt}}
\approx \exp ( WL_{opt}/\Delta) $. This is an exponential ``gain''
in ${\cal P}(\ln T, {\cal L})$. The ``loss'' due to suppression
of the gap, similarly to Eq.~(\ref{exponent}), can be
expressed as $\sim \exp\left[  - L_{opt} (\delta \epsilon
/\Delta )^2\right]$. The relation between $\delta \epsilon $ and
$W$ can be established from Eq.~(\ref{d_e_0}). Indeed, all the terms in
numerator are of the same sign and of the order of $W$, while the
corresponding terms in denominator are also sign-preserving and
$\sim \Delta$. Thus we have $\delta \epsilon \sim (\pi/n) (W/\Delta )$.
Finally, the product
\begin{eqnarray}
& & \exp \left[ L_{opt} \left(\frac{W}{\Delta}\right)\right]
\exp \left[ -  L_{opt} \left(
\frac{\delta \epsilon }{\Delta } \right)^2     \right]  =  \nonumber \\
& & ~~~ = \exp \left\{  L_{opt} \left[
\left(\frac{W}{\Delta}\right) - \left(
\frac{\pi W }{n \Delta ^2 } \right)^2
 \right]  \right\}
\label{W}
\end{eqnarray}
of the gain and loss has a maximum at $W_{opt} \approx n^2 \Delta ^3$.
We see, that the allowance is relatively
small, since $W_{opt}/\Delta = n^2 \Delta ^2 =
(1/n) (E_t/\epsilon_n)^{3/2} \ll 1$. This suggest
that the allowance does not change the result
Eq.~(\ref{small1}), since the correction to
$\ln {\cal P} (\ln T)$ due to allowance, $W_{opt}$,
amounts to a small portion
$\sim W_{opt}/\Delta \ll 1$ of the main exponent
Eq.~(\ref{small1}).

Strictly speaking, the optimal value of the
allowance, $W$, is well defined if the exponent
in Eq.~(\ref{W}) is much bigger than one.
Upon substituting $W_{opt}$ into Eq.~(\ref{W}), we obtain
$\delta |\ln {\cal P}| \sim n^2 \Delta^2 L_{opt} \approx
n \Delta |\ln T| $. From here we conclude, that
for large enough $|\ln T|> 1/(n\Delta )$,
the allowance, $W$, indeed leads to $\delta |\ln {\cal P}|
\gg 1$. For smaller $|\ln T|\lesssim 1/(n\Delta )$
the gain does not play a role, so that
the allowance, $W$, is determined exclusively by the
second term in the exponent in Eq.~(\ref{W}). This
term falls off at characteristic $W = n \Delta ^2
/L_{opt}^{1/2}$.
This allowance is bigger than
$W_{opt}$, since $W/W_{opt} \approx (n\Delta |\ln T|)^{-1/2}>1$,
but still smaller than $\Delta$. The latter can be seen if
we rewrite  $W/\Delta = n \Delta  /L_{opt}^{1/2}$ in the form
$W/\Delta \approx \left( W_{opt}/\Delta \right)^{3/4}
|\ln T|^{-1/2}$, which is the product of two small numbers.
The fact that for small $|\ln T|\lesssim 1/(n\Delta )$
the allowance increases with $|\ln T|$ can be
interpreted qualitatively as follows.
The smaller is $|\ln T|$,
the less effort is required to create a disorder
configuration with low transmission $T$.

The enhancement of the portion of the loose mirrors
due to allowance, $\kappa $, is equal to
$\left( 2^{\kappa } \right)^{L_{opt}/n} =
\exp\left( \kappa (L/n) \ln 2\right)  $.
The loss can be estimated analogously to the
case (i). Namely, due to the sites with wrong
sign of on-site energies the ratio
$\sum_{m=1}^{n} V_m^{(i)} \sin \left( \frac{\pi m}{n}\right) /
 \sum_{m=1}^{n} V_m^{(i)} \cos \left( \frac{\pi m}{n}\right) $
in Eq.~(\ref{d_e_0}) is $\sim \kappa$. This yields the
estimate $\delta \epsilon \sim (\pi/n) \kappa$, so that,
analogously to (i), the product of gain and loss
can be written as
\begin{equation}
\label{kappa}
\exp \left\{  L_{opt} \left[   \kappa -
\left( \frac{\pi \kappa  }{n \Delta  } \right)^2
\right]  \right\}.
\end{equation}
This product is maximal for $\kappa _{opt} = n^2 \Delta ^2$.
We see that $\kappa _{opt} \approx W_{opt} /\Delta $, and
thus is small, as discussed above.

The fact that the optimal
allowances $W_{opt}/\Delta $ and $\kappa _{opt}$ are small
justifies that loose mirrors are well-defined entities.

\vspace{3mm}

\centerline{\bf V. CONTINUOUS LIMIT}

\vspace{4mm}

In this Section we establish the relation between
the above consideration on the lattice and the results of
Refs.\onlinecite{muzykantskii,smolyarenko}, obtained within
the continuous approach.
To establish this relation, we restore the lattice constant, $a$, in the
dispersion law, i.e. $\epsilon (k) a^2 = 4 \sin^2\left(ka/2\right)$,
where $k$ is the momentum. For lattice constant $a=1$
the dimensionless parameter  that separates
the regimes of weakly disturbed  [Eqs. (\ref{small}), (\ref{large})] and loose
[Eqs. (\ref{small1}), (\ref{large1})] mirrors was equal to
$n \Delta^{2/5}$. To incorporate the arbitrary lattice constant,
it is convenient to first express this parameter through the
conductance $G$ for $a=1$. From the relation $G_n = n^{-3}\Delta ^{-2}$
we find $\Delta ^{2/5} = n^{-3/5} G_n^{-1/5}$. Thus,
$n \Delta^{2/5} = n^{2/5}/G_n^{1/5}$. For arbitrary $a$,
the number $n$ should be replaced by $(ka)^{-1}$, while $G_n$
should be replaced by $G(k) = \left( \epsilon (k)/E_t \right)^{3/2}$,
where $E_t$ is the upper boundary of the tail states. For $a=1$
this boundary is expressed through $\Delta $ as
$E_t = \Delta ^{4/3}$. Thus, the parameter $n \Delta^{2/5}$
for arbitrary $a$ transforms into
$G(k)^{-1/5}\left(ka\right)^{-2/5}$. It is seen that
this parameter contains the lattice constant and in the
white noise limit $a\rightarrow 0$, considered in
Refs.\onlinecite{muzykantskii,smolyarenko},
it is much bigger than 1, {\em which corresponds to
the regime of loose mirrors}.
We thus conclude,  that for small $a$ the distribution function
${\cal P} \left(\ln T\right)$ is given by
Eqs. (\ref{small1},\ref{large1}). With $a$ restored, these
expressions take the form
\end{multicols}
\widetext
\vspace*{-1mm} \noindent  \hrulefill \hspace*{3.6truein}
\begin{eqnarray}
\left\vert \mbox{\Large $\ln$ } \! {\cal P}\left(\ln T, {\cal L}\right)
       \right\vert = & & \beta_{eff}
\left[  \frac{2 \pi^3 G(k)}{ka } \right]^{1/2}
                                       \vert \ln T \vert
,~~~~~~~~~~
\frac{\beta_0^{1/2}}{\pi^{3/2}}
\left[\frac{k{\cal L}}{G^{1/2}(ka)^{1/2}} \right] >
         \vert \ln T \vert >  \frac{ k {\cal L}}{\pi G}
\label{small2}
    \\
\left\vert \mbox{\Large $\ln$ } \! {\cal P}\left(\ln T, {\cal L}\right)
       \right\vert = & &
\left[ \frac{\pi^3 G}{\beta_2 {\cal L} \sqrt{\epsilon } } \right]
  \ln ^2 T -
\left[  \frac{2 \pi^3 \beta_1^2 G}{\beta_2^2 a \sqrt{\epsilon }}
\right]^{1/2} \!\!\! | \ln T| -
\left( \frac{\beta_0}{\beta_2} \right) \frac{{\cal L}}{a}
     ,~~~~
 \vert \ln T \vert >
\frac{\beta_0^{1/2}}{\pi^{3/2}}
\left[\frac{k{\cal L}}{G^{1/2}(ka)^{1/2}} \right]
 \!\! .
\label{large2}
\end{eqnarray}

\hspace*{3.6truein}\noindent  \hrulefill

\begin{multicols}{2}

The result of Refs.\onlinecite{muzykantskii,smolyarenko}
correspond to the first term of Eq. (\ref{large2}). We see, however,
that this result, obtained  neglecting the prefactor, does not
survive the white-noise limit $a \rightarrow 0$. Formally,
taking the prefactor into account, shifts the domain of applicability
of the log-normal distribution to very small transmission coefficients
$\vert \ln T \vert > \left(  \beta_0  k {\cal L}^2 /\pi ^3 G a \right)^{1/2}$, so that
this domain vanishes when $a \rightarrow 0$.
Physically, the result of Refs.\onlinecite{muzykantskii,smolyarenko},
does not apply in the white-noise limit due to the huge
phase fluctuations, that are allowed for small $a$. These fluctuations
forbid the formation of a Bragg mirror with a weakly
distorted sinusoidal shape of Fig.~3a.
Instead, the relevant fluctuations have the form of loose
mirrors shown in  Fig.~3b,
where only the positions of sign changes are adjusted
to the de Broglie wave length, $2\pi/k_F$, of the electron.
In this regime the tail of
  ${\cal P}\left(\ln T\right)$ is described by a simple
exponent Eq. (\ref{small2}) with a non-universal coefficient
$\propto a^{-2}$. Clearly, in the ``continuous language'',
the lattice constant $a$ should be identified with
the smallest scale in the problem, namely,
the {\em correlation length} of the random potential.
Thus, we arrive at the conclusion that the correlation
length determines the coefficient in front of
$| \ln T |$ in the leading term of ${\cal P}(\ln T)$.
In terms of the correlation length and dimensionless conductance,
the portion of the sample occupied by the loose mirror
is given by
$\left\vert\ln T\right\vert  \left(  Ga /\beta_0 k {\cal L}^2 \right)^{1/2}$.

Finally, we establish the energy interval, where the loose
mirrors, and thus Eqs.~(\ref{small2}), (\ref{large2}),
determine the far tail of ${\cal P} (\ln T)$. For this
purpose, we equate the parameter $G(k)^{-1/5}\left(ka\right)^{-2/5}$
to 1 and express the energy $E_B$ in Fig.~4 through the correlation
length $a$ as $E_B \approx E_t /(E_t a^2 )^{2/5}$. This yields
the sought interval
\begin{equation}
\label{interval}
E_t < \epsilon < \frac{ E_t }{\left( E_t a^2 \right)^{2/5}}.
\end{equation}
Recall that $E_t$, the position of the boundary of the
tail states, does not depend on $a$. This is valid for
small $a$, such that $(E_t a^2) \ll 1$, i.e. the interval
 (\ref{interval}) is broad.
For Eq. (\ref{interval}) to apply, we should require that at the
upper boundary of the interval (\ref{interval}) the
corresponding momentum, $k_B$, is much smaller than the inverse
correlation length. It is easy to see that this is indeed
the case, since $ k_B a  = a E_B^{1/2} = (E_t a^2)^{3/10}\ll 1 $.

Fig. 7 illustrates the main qualitative outcome of our consideration.
Namely, for a short-range potential with a correlation radius, $a \ll k_F^{-1}$,
the low-$T$ disorder configurations for energies within the interval Eq. (\ref{interval})
have the shape of loose mirrors depicted schematically in this figure.

\begin{figure}
\narrowtext
\centerline{
\epsfxsize=3.2in
\epsfbox{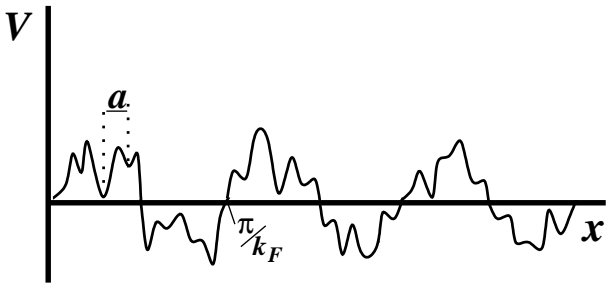}
\protect\vspace*{0.1in}}
\protect\caption[sample]
{\sloppy{``Continuous'' realization of a loose mirror for
short-range  disorder with correlation
radius
$a \ll k_F^{-1}$.
}}
\label{f7}
\end{figure}

\vspace{1mm}

\centerline{\bf VI. CONCLUSION}

\vspace{3mm}

In conclusion, let us address the relation between our results
 and the analytical
results in 1D, predicting the shape of
${\cal P}(\ln T)$.
Neither of the ``exact'' techniques\cite{berezinskii,gorkov}
allows to pinpoint the actual disorder configuration,
responsible for low-$T$ values. Although they are believed to
be exact, each of these techniques contain a step at which mirror-like
configurations are lost. Let us illustrate this point
using the Berezinskii technique\cite{berezinskii} as an example.
In Fig. 8 a three-impurity scattering configuration, employed in
Ref. \onlinecite{berezinskii} (see also\cite{anderson}) to make the case
for complete
localization in 1D,
 is depicted.
As was explained by Berezinskii\cite{berezinskii}, the key ingredient
of the technique Ref.~\onlinecite{berezinskii} is the observation that the
scattering paths I and II correspond to the {\em same} accumulated phase
\begin{equation}
\label{phase}
\phi = 2 k_F (x_2 + x_3 - 2 x_1),
\end{equation}
and, thus, interfere
constructively for {\em any } $\phi $. However, within the ``exact''
technique, the value of $\phi $ is assumed to be random, and the
averaging over $\phi $ is performed. Similar procedure
is a key element of the  technique Ref.\onlinecite{gorkov}. Calculating
the higher-order diagrams in Ref.~\onlinecite{altshuler89} takes
into account increasingly large number of possibilities of
constructive interference  of different paths, all of which are of the
type Fig.~8
(correspond to the same accumulated {\em random} phase). However, each
step involves averaging over this phase.
In contrast, the Bragg-like configurations are those sparse
realizations, for which the phase accumulated upon
traversing the period, $\pi/ k_F$, first forwards, and then
backwards, is {\em not random}, but close to $2\pi$.
Thus, in our opinion,
the complete coincidence of the estimate for ${\cal P} (\ln T)$
based on
the Bragg mirrors and of the  result\cite{altshuler89} is accidental.

\begin{figure}
\narrowtext
\centerline{
\epsfxsize=2.3in
\epsfbox{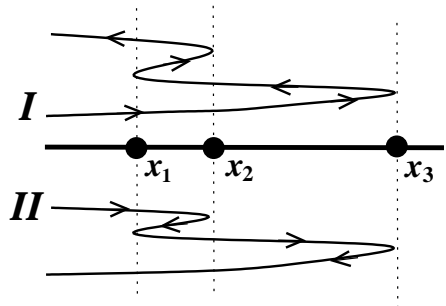}
\protect\vspace*{0.1in}}
\protect\caption[sample]
{\sloppy{Illustration of the simplest building block of the
Berezinskii technique Ref.~\onlinecite{berezinskii}. In course of moving along
the trajectories, I and II,
an electron accumulates the same {\em random} phase
$\phi=2k_F(x_2+x_3-2x_1)$.
}}
\label{f8}
\end{figure}

Finally, let us briefly formulate the main message of the present paper.
Creating low-$T$ disorder configuration in 1D demands this configuration
to possess a long-range order, adjusted to the wave vector, $k_F$.
Ideal configurations with such a long-range order, are of the
Bragg-mirror type. However, they
are very ``costly'' to maintain over a large distance. This is
due to the phase fluctuations,
that tend to violate the Bragg condition. These fluctuations are
not captured at the stage of calculating the saddle-point.
 They show up at the next stage, i.e. calculating the prefactor.
We have demonstrated that loose mirrors, illustrated in Fig. 7,
in which the
long-range order is present, but relaxed, are much ``cheaper'' to
create than the
Bragg mirrors. On the other hand, as follows from the analysis that
we have performed,
a loose mirror, shown
in Fig. 7, still constitutes an efficient low-$T$ disorder configuration.
The smaller
is the correlation radius, $a$, of the disorder, the wider is the
energy interval
within which loose mirrors dominate the low-$T$ tail of the
transmission distribution.

Lastly, we are not aware of any numerical work in which the
tail of ${\cal P}(\ln T)$ was studied close to (but well above) the band edge.
Recent simulations are mostly focused at the body of the distribution
both in 1D\cite{deych,uski01} and in 2D\cite{slevin01,slevin04}, and are aimed
at testing the scaling hypothesis. With regard to the tail of the
transmission distribution, the related characteristics, namely,
the density of anomalously localized states, was studied numerically
only in two\cite{uski00} and three\cite{uski02,nikolic'01,nikolic02,nikolic'02}
dimensions.





\vspace{3mm}

\centerline{\bf ACKNOWLEDGEMENTS}

\vspace{3mm}
We acknowledge the hospitality of the Max-Plank Institute
for Complex Systems, where the part of this work was
completed, and the support of
the Petroleum Research Fund under Grant No. 37890-AC6.
We also acknowledge stimulating  discussions with
B. Shapiro.

\end{multicols}


\begin{references}


\large







\bibitem{deych}L. I. Deych, A. A. Lisyansky, and B. L. Altshuler,
Phys. Rev. Lett. {\bf 84},  2678 (2000); Phys. Rev. B {\bf 64},
224202 (2001), and references therein.

\bibitem{schomerus} H. Schomerus, M. Titov, P. W. Brouwer,
and C. W. J. Beenakker, Phys. Rev. B {\bf 65}, 121101(R) (2002),
and references therein.




\bibitem{berezinskii} V. L. Berezinskii, Zh. Eksp. Teor. Fiz.
{\bf 65}, 1251 (1973) [Sov. Phys. JETP {\bf 38}, 620 (1974)].


\bibitem{izrailev} F. M. Izrailev and A. A. Krokhin,
 Phys. Rev. Lett. {\bf 82}, 4062 (1999).


\bibitem{altshuler} B. L. Altshuler, V. E. Kravtsov, and I. V. Lerner,
in {\em Mesoscopic Phenomena in Solids}, edited by B. L. Altshuler,
P. A. Lee, and R. A. Webb (North-Holland, Amsterdam, 1991).

\bibitem{khmelnitskii} B. A. Muzykantskii and D. E. Khmelnitskii,
Phys. Rev. B {\bf 51}, 5480 (1995).

\bibitem{falko} V. I. Fal'ko and K. B. Efetov, Phys. Rev.
B {\bf 52}, 17 413 (1995).


\bibitem{mirlin}  A. D. Mirlin, Phys. Rep. {\bf 326}, 259 (2000),
 and references therein.



\bibitem{muzykantskii} B. A. Muzykantskii and  D. E. Khmelnitskii,
preprint cond-mat/9601045.

\bibitem{smolyarenko}I. E. Smolyarenko and B. L. Altshuler,
Phys. Rev. B {\bf 55}, 10451 (1997).

\bibitem{altshuler89} B. L. Altshuler and V. N. Prigodin,
Zh. Eksp. Teor. Fiz. {\bf 95}, 348 (1989)
[Sov. Phys. JETP {\bf 68}, 198 (1989)].

\bibitem{halperin} B. I. Halperin and M. Lax, Phys. Rev. {\bf 148},
722 (1966).

\bibitem{zittartz} J. Zittartz and J. S. Langer, Phys. Rev. {\bf 148},
 741 (1966).





\bibitem{halperin67} B. I. Halperin and M. Lax, Phys. Rev. {\bf 153},
802 (1967).

\bibitem{houghton79} see, {\em e.g.}, A. Houghton and L. Sch\"{a}fer, J. Phys. A
{\bf 12}, 1309 (1979).

\bibitem{brezin} E. Br\'{e}zin and G.  Parisi,
J. Phys. C: Solid State Phys. {\bf 13}, L307 (1980).

\bibitem{cardy}J. L. Cardy, J. Phys. C: Solid State Phys. {\bf 11},
L321 (1978).

\bibitem{we} V. M. Apalkov, M. E. Raikh, and B. Shapiro,
Phys. Rev. Lett. {\bf 92}, 066601 (2004).

\bibitem{gorkov} V. L. Berezinskii and L. P. Gor'kov,
Zh. Eksp. Teor. Fiz. {\bf 77}, 2489 (1979)
[Sov. Phys. JETP {\bf 50}, 1209 (1979)].

\bibitem{anderson}P. W. Anderson, D. J. Thouless, E. Abrahams,
and D. S. Fisher, Phys. Rev. B {\bf 22}, 3519 (1980).

\bibitem{uski01}V. Uski, B. Mehlig,
and M. Schreiber, Phys. Rev. B {\bf 63}, 241101(R) (2001).

\bibitem{slevin01}K. Slevin, P. Marko\v{s}, and T. Ohtsuki, Phys. Rev. Lett. {\bf 86},
3594 (2001); Phys. Rev. B {\bf 67}, 155106 (2003).

\bibitem{slevin04} K. Slevin, Y. Asada, and L. I. Deych, preprint cond-mat/0404530.



\bibitem{uski00} V. Uski,
 B. Mehlig, R. A. R\"{o}mer, and M. Schreiber,
Phys. Rev. B {\bf 62}, R7699 (2000).



\bibitem{uski02}V. Uski, B. Mehlig, and M. Schreiber,
Phys. Rev. B {\bf 66}, 233104 (2002).


\bibitem{nikolic'01}B. K. Nikoli\'{c}, Phys. Rev. B {\bf 64},
014203 (2001).


\bibitem{nikolic02}B. K. Nikoli\'{c}, Phys. Rev. B {\bf 65},
012201 (2002).

\bibitem{nikolic'02}B. K. Nikoli\'{c}, V. Z. Cerovski,
Eur. Phys. J. B {\bf 30}, 227 (2002).











\end{references}
\end{document}